Thermophysical assessment on the feasibility of basal melting in the south polar region of Mars

Lujendra Ojha[1], Jacob Buffo[2], Baptiste Journaux[3]

[1]Department of Earth and Planetary Sciences, Rutgers University, Piscataway, NJ, USA.
[2]Thayer School of Engineering. Dartmouth College. Hanover, NH, USA.
[3]Department of Earth and Space Science, University of Washington, Seattle, WA, USA.

Corresponding Author: (luju.ojha@rutgers.edu)

**Key Points:**

- The feasibility of the potential existence of a subglacial lake on the south pole of Mars is investigated using thermal models.

- Subglacial lakes cannot be reconciled with current constraints on Martian heat flow and the bulk composition of the south pole.

- Exceptional circumstances would be necessary to form subglacial lakes in the Martian south pole.

Abstract

Bright basal reflectors in radargram from the Mars Advanced Radar for Subsurface and Ionosphere Sounding (MARSIS) of the Martian south polar layered deposits (SPLD) have been interpreted to be evidence of subglacial lakes. However, this interpretation is difficult to reconcile with the low Martian geothermal heat flow and the frigid surface temperature at the south pole. We conduct a comprehensive thermophysical evolution modeling of the SPLD and show that subglacial lakes may only form under exceptional circumstances. Subglacial lakes may form if the SPLD contains more than 60 % dust volumetrically or extremely porous ice (>30 %), which is unlikely. A thick (>100 m) layer of dirty ice (>90% dust) at the base of the SPLD may also enable basal melting, resembling a sludge instead of a lake. Other scenarios enabling subglacial lakes in the SPLD are equally unlikely, such as recent magmatic intrusions at shallow depths.

Plain Language Summary

Radar data from the base of the Martian south polar cap have been interpreted by some to be evidence of a liquid water lake. However, this interpretation is hard to reconcile with the observation that the surface temperature of the Martian south pole is frigid (-108° C). Recently, it has been proposed that the base of the Martian south polar cap contains extremely salty ice that can melt at -74° C; thus, despite the freezing conditions at the Martian south pole, liquid water may form at the base. In this paper, we create various numerical models to understand how heat is transmitted through the southern polar cap of Mars and if a salty liquid water lake can form as a result. We show that exceptional requirements are necessary for heat to be transferred efficiently to form liquid water at the base of the south polar cap. The conditions required for liquid water to form at the base of the Martian southern polar cap are at stark odds with what we currently know about the composition of the polar cap. In summary, numerical model results favor alternative, dry interpretations of the radar data.

1 Introduction

The south polar region of Mars is primarily composed of $H_2O$ ice in combination with minor amounts of $CO_2$ ice and dust. The SPLD's gravity-derived bulk density estimate of 1100 – 1300 kg m$^{-3}$ suggests the presence of dust content of at least 9% by volume (up to 20 %) if the $CO_2$ concentration is assumed to be negligible [*Broquet et al.*, 2021; *Li et al.*, 2012; *Wieczorek*, 2008; *Zuber et al.*, 2007]. Observations by the Shallow Radar instrument on Mars Reconnaissance Orbiter also reveal small volumes of buried $CO_2$ ice (< 1%) within the south polar layered deposits [*Bierson et al.*, 2016; *Phillips et al.*, 2011]. These impurities provide a glimpse into the million-year variation in the orbital parameters of Mars leading to the differential deposition of dust and $CO_2$ in the south polar cap. Furthermore, the presence of impurities like dust, $CO_2$, and other possible clathrates can affect the thermal structure of the south polar ice cap, possibly raising the basal temperature to enable basal melting [*Greve and Mahajan*, 2005].

Recently, bright basal reflectors in the radargram of Ultimi Scopuli have been interpreted to be evidence of a subglacial lake system [*Lauro et al.*, 2022; *Lauro et al.*, 2021; *Orosei et al.*, 2018], although this interpretation has been challenged by multiple authors [*Lalich et al.*, 2022; *Ojha et al.*, 2021; *Smith et al.*, 2021]. While subglacial lakes are common in various areas of Earth [*Livingstone et al.*, 2022], the main factor that hinders the formation of subglacial lakes on Mars is the frigid temperature at the south pole (<165 K) [*Sori and Bramson*, 2019] and expected low geothermal heat flow on present-day Mars [*Broquet et al.*, 2021; *Ojha et al.*, 2021; *Ojha et al.*, 2019; *Parro et al.*, 2017; *Plesa et al.*, 2018]. Therefore, augmentation of the basal temperature by exceptionally high heat flow from a recent magmatic intrusion would be necessary to melt pure water ice at the Martian south pole [*Sori and Bramson*, 2019]. Alternatively, if one were to suppose that the base of the SPLD is composed of saline ice, then the upper limit on the temperature required for melting can be relaxed to a much lower value.

Although several recent papers have provided alternative interpretations of the bright radar reflections, including $CO_2 - H_2O$ layer boundaries [*Lalich et al.*, 2022], saline ice [*Bierson et al.*, 2021], or smectites [*Smith et al.*, 2021], the debate has persisted. Lauro et al. (2022) asserted that the subglacial lakes at the Martian south pole are composed of perchlorate brines with a eutectic temperature of ~200 K. Further, Lauro et al. (2022) claim that basal melting can be achieved in the SPLD without a recent magmatic intrusion and with reasonable assumptions about the presence of porous ice and impurities like dust and $CO_2$ within the SPLD. Here, we assume that the interpretation of the subglacial lake is correct. With liquid water (pure or briny) at the base of the SPLD as an a priori condition, we conduct a comprehensive thermophysical evolution modeling exercise to ascertain the conditions that can enable basal melting at the south pole of present-day Mars.

## 2 Data and Methods

### 2.1 1-dimensional thermal models to assess the feasibility of basal melting due to the bulk impurities within the SPLD

We set up a 1-D thermophysical evolution model of the SPLD in COMSOL that simulates coupled heat transfer and phase change. COMSOL utilizes the apparent heat capacity formulation to provide an implicit method of capturing the phase change interface by solving for both phases and the heat transfer equation with effective material properties via:

$$\rho C_{eq} \frac{\partial T}{\partial t} + \nabla \cdot \left( -k_{eq} \nabla T \right) = 0;$$

*(eqn 1)*

where $\rho$ (kg m$^{-3}$) is the SPLD density, $C_{eq}$ (J kg$^{-1}$ K$^{-1}$) is the effective heat capacity, $k_{eq}$ (W m$^{-1}$ K$^{-1}$) is the effective thermal conductivity, $T$ (K) is temperature, and $\nabla$ is the del operator. The latent heat of phase change ($L$) is taken into account by modifying the effective heat capacity (i.e., the apparent heat capacity formulation) as:

$$c_p = \frac{1}{\rho} \left( \theta_1 \rho_1 c_{p,1} + \theta_2 \rho_2 c_{p2} \right) + L_{1 \rightarrow 2} \frac{\partial \alpha_m}{\partial T};$$

*(eqn 2)*

where $c_p$ is the bulk apparent heat capacity, $c_{p,1}$ is the specific heat capacity of phase 1, and $c_{p,2}$ is the specific heat capacity of phase 2. Here, $L_{1 \rightarrow 2}$ is the latent heat of melting which we vary with the melting temperature of the ice (see Section 2.2). The parameter $\alpha_m$ is the mass fraction given by:

$$\alpha_m = \frac{1}{2} \frac{\theta_2 \rho_2 - \theta_1 \rho_1}{\theta_1 \rho_1 + \theta_2 \rho_2}.$$

*(eqn 3)*

In equation 2 and 3, $\theta_1$ and $\theta_2$ are the mass fractions of phase 1 and phase 2 (i.e., $\theta_1 + \theta_2 = 1$). As phase change occurs, the value of various thermophysical parameters are computed as a volumetric average of $H_2O$-ice ($\theta_1$) and liquid water ($\theta_2$) given by:

$$\rho = \theta_1\rho_1 + \theta_2\rho_{2;}$$

and

$$k = \theta_1 k_1 + \theta_2 k_{2.}$$

(eqn 4)

The thermal evolution models are run for a minimum of 2 million years (Text S1). We set the surface temperature at the top of the ice sheet to 165 K [*Sori and Bramson*, 2019]. At the base of the SPLD, a net-inward flowing heat flux boundary condition is imposed, which we vary between 5 - 60 mW m². The accuracy of the apparent heat capacity formulation in capturing phase change is verified by comparing the output of the COMSOL simulation to the analytical solution of the Stefan problem (Text S2; Fig. S1).

2.2 Thermophysical Parameters of water-ice and mixtures:

We use SeaFreeze [*Journaux et al.*, 2020] to estimate the temperature and pressure dependent thermophysical parameters of pure water-ice such as $c_p$ and $\rho$ (Fig. S2). The latent heat of melting of ice is also temperature dependent which we estimate using SeaFreeze (Fig. S3). The thermal conductivity of pure water-ice ($k_{ice}$) is temperature dependent and can be estimated using the following equation [*Petrenko and Whitworth*, 1999]:

$$k_{ice}(z) = \frac{651}{T_{(z)}}.$$

(eqn 5)

Once the temperature-dependent thermophysical parameters of pure ice are computed, we use mixing models to estimate the thermophysical parameters of water-ice and dust mixtures. We estimate the bulk thermal conductivity of the dusty ice ($K_{bulk}$) using the heterogenous two-component system mixing model [*Hamilton and Crosser*, 1962]:

$$K_{bulk}(z) = K_{ice}(z)\left[\frac{K_{dust}+[n-1]K_{ice(z)}-[n-1]V_{dust}[K_{ice(z)}-K_{dust}]}{K_{dust}+[n-1]K_{ice(z)}+V_{dust}[K_{ice(z)}-K_{dust}]}\right].$$

(eqn 6)

The parameter n in equation 6 is a dimensionless, empirical shape factor set to 3. This is a valid approach as long as the thermal conductivity of the discontinuous phase (i.e., dust) is less than a factor of 100 different than the continuous phase (i.e., $H_2O$ ice) [*Hamilton and Crosser*, 1962]. $V_{dust}$ represents the volume of dust present in the SPLD ranging between 0 and 1. Other thermophysical parameters, such as $c_p$ and $\rho$ of the SPLD, are computed as volumetric averages of dust and ice (Fig. S2; Table S1). The thermal conductivity of Martian dust is set to 0.039 W m$^{-1}$ K$^{-1}$, as measured by InSight [*Grott et al.*, 2021]. We also consider a slightly lower value of 0.015 W m$^{-1}$ K$^{-1}$ based on laboratory measurements of the thermal conductivity of Martian dust simulants [*Yu et al.*, 2022]. The mixing models are coupled with the thermal evolution model because as the thermal model evolves with time, the temperature at any particular depth (z) can change, affecting the thermophysical parameters of the water-ice and the bulk SPLD.

2.3 Thermophysical Parameters of porous water-ice:

Porosity within the SPLD can also notably affect the thermal structure. We utilize a parametereized model that relates the porosity and temperature to the thermal conductivity of snow, firn, and porous ice $k_{\text{snow-firn}}(\rho, T)$ given by [*Calonne et al.*, 2019]:

$$k_{\text{snow-firn}}(\rho, T) = (1 - \theta) \frac{k_{ice}(T) k_{air}(T)}{k_{ice}^{\text{ref}} k_{air}^{\text{ref}}} k_{\text{snow}}^{\text{ref}}(\rho) + \theta \frac{k_{ice}(T)}{k_{ice}^{\text{ref}}} k_{\text{firn}}^{\text{ref}}(\rho)$$

*(eqn 7)*

where $k_{ice}(T)$ is the temperature dependent thermal conductivity of water-ice given by equation 5, $k_{air}(T)$ is the temperature dependent thermal conductivity of air. Based on [*Calonne et al.*, 2019], we set the $k_{ice}^{\text{ref}}$ and $k_{air}^{\text{ref}}$ to 2.107 W m$^{-1}$ K$^{-1}$ and 0.024 W m$^{-1}$ K$^{-1}$ respectively. Because the thermal conductivity of air does not notably change with temperature, we set $k_{air}(T)$ to be equal to $k_{air}^{\text{ref}}$. Other parameters in equation 7 are parameterized as the follows:

$$\theta = \frac{1}{\left(1 + \exp\left(-2a(\rho - \rho_{\text{transition}})\right)\right)};$$
$$k_{\text{firn}}^{\text{ref}}(\rho) = 2.107 + 0.003618(\rho - \rho_{ice});$$
$$k_{\text{snow}}^{\text{ref}}(\rho) = k_{\text{cal.}} = 0.024 - 1.23 \times 10^{-4}\rho + 2.5 \times 10^{-6}\rho^2;$$

*(eqn 8)*

where $\rho_{ice}$ is set to 917 kg m$^{-3}$ and $\rho_{\text{transition}}$ is set to 450 kg m$^{-3}$. The above set of equations are solved as a function of temperature and density of ice ($\rho$). Porosity ($\emptyset$) is computed as follows:

$$\emptyset = \left(1 - \frac{\rho_{ice}}{\rho_{SPLD}}\right).$$

*(eqn 9)*

## 2.4 2-dimensional magmatic instrusion model

Previously, Sori and Bramson (2019) found that augmentation of the geothermal heat flux by at least 72 mW m$^{-2}$ is required for basal melting under the most favorable compositional considerations. However, the size of the putative lake was not considered in that work. Here, we explore the possible geometry, thermal budget, and depth of the magmatic intrusion required to form a subglacial lake approximately 20 - 30 km in width [*Lauro et al.*, 2021]. Another significant difference between our magmatic intrusion model and that of Sori and Bramson. (2019) is that we prescribe the magmatic intrusion with an initial temperature value instead of a heat flux. Furthermore, since relatively cold country rocks would surround any recent intrusion on Mars, we allow the temperature of the magmatic intrusion to cool over the timescale of the simulation (Text. S3; Fig. S4). Incorporating a time-dependent cooling intrusion also allows us to assess the longevity of any brines produced by an intrusion.

We consider a 50-km thick Martian crust superposed by a 1.5 km thick ice sheet resembling the SPLD. Initially, we model the steady-state temperature profile of the crust by assuming a surface heat flow of 30 mW m$^{-2}$ (Text S3; Fig. S5). A magmatic intrusion of variable geometry and depth is then introduced within the crust to assess its overall impact on the feasibility of basal melting. As melting occurs within the SPLD, convection can play a dominant role; thus, we couple the heat transfer module within COMSOL to the laminar flow package Multiphysics to account for possible convection within the melt layer. As the density between the two phases is different, we solve the "Weakly Compressible" form of the fluid flow equations and use a moving mesh boundary condition to account for the overall volume change of the ice/water layer. The viscosity of the liquid water is set to 0.018 Pa. s, whereas the viscosity of the $H_2O$-ice is set to $10^5$ Pa. s. A more detailed explanation of the model setup is provided in Text S3. Finally, we set the initial temperature of the intrusion to 1500 K, a relatively high value on par with the temperature of some of the hottest magmas preserved on Earth [*Arndt et al.*, 2008].

## 3 Results and discussion

### 3.1. Can bulk impurities within the SPLD enable basal melting?

We first consider a scenario in which we assess if the bulk impurities and porosity present within the SPLD can enable basal melting. We vary $T_s$ between 165 – 170 K and solve for the $k_{bulk}$ of SPLD that would allow basal melting as a function of $q_s$. The $k_{bulk}$ necessary for basal melting increases with $q_s$, and variation in $T_s$ do not notably impact the results (Fig. 1 a). It is unlikely that the regional heat flow in the south polar region exceeds 30 mW m$^{-2}$ [*Broquet et al.*, 2021; *Ojha et al.*, 2021; *Ojha et al.*, 2019; *Parro et al.*, 2017; *Plesa et al.*, 2018]; thus, $k_{bulk} \leq 1.5$ W m$^{-1}$ K$^{-1}$ is necessary for the SPLD to undergo basal melting (Fig. 1 a).

We solve for the combination of $H_2O$-ice, porosity, and dust content that can yield $k_{bulk} \leq 1.5$ W m$^{-1}$ K$^{-1}$. We initially assume the bulk of the SPLD to be pore-free and solve for the volumetric dust content required for basal melting using the mixing relation for $H_2O$-ice and dust (eqn 6). Dust content above 60 %, regardless of the two $k_{dust}$ value considere here, would be required for basal melting to occur with a surface heat flow of 30 mW m$^{-2}$ (Fig. 1 b). The bulk impurities necessary for basal melting from our thermal simulation can be compared to the bulk composition of the SPLD inferred from gravity [*Broquet and Wieczorek*, 2019; *Li et al.*, 2012; *Wieczorek*, 2008; *Zuber et al.*, 2007] and radar data [*Lauro et al.*, 2022]. While there are uncertainties in the possible volume of dust present within the SPLD, it is unlikely to exceed 20 %, significantly lower than the bulk volume required for basal melting in the SPLD (Fig. 1 c). The presence of $CO_2$ ice within the SPLD can also aid in the feasibility of basal melting due to its relatively low thermal conductivity of 0.02 W m$^{-1}$ K$^{-1}$. However, there is no evidence in the radar data to suggest hidden voluminous (> 1%) deposits of $CO_2$ within the SPLD [*Bierson et al.*, 2016; *Phillips et al.*, 2011].

We now consider if pore spaces within the SPLD can enable basal melting. Using the parameterized mixing model of Calonne et al. (2019), we solve for the $k_{bulk}$ as a function of the temperature and porosity of ice (Fig. 1 d). Regardless of the temperature, porosity in excess of 0.3 would be necessary for the $k_{bulk}$ to be lower than 1.5 W m$^{-1}$ K$^{-1}$. The required porosity for basal melting can be compared with the bulk density constraint of the SPLD derived from gravity data. Considering $H_2O$-ice ($\rho_{ice} = 930\ kg\ m^{-3}$), dust ($\rho_{dust} = 3000\ kg\ m^{-3}$), and air ($\rho_{air} = 0.02\ kg\ m^{-3}$), to be the main components of the SPLD, bulk porosity $\leq 0.2$ can be present within the SPLD and still satisfy the gravity-derived bulk density constraint (Fig. 1 c). If a slightly lower $\rho_{dust}$ of 2500 kg m$^{-3}$ is considered, then the bulk porosity of the SPLD cannot exceed 0.12. In either case, the required porosity of 0.3 for basal melting cannot be reconciled with the gravity-derived density of the SPLD. Thus, it is unlikely that the bulk impurities or porosity within the SPLD can enable basal melting.

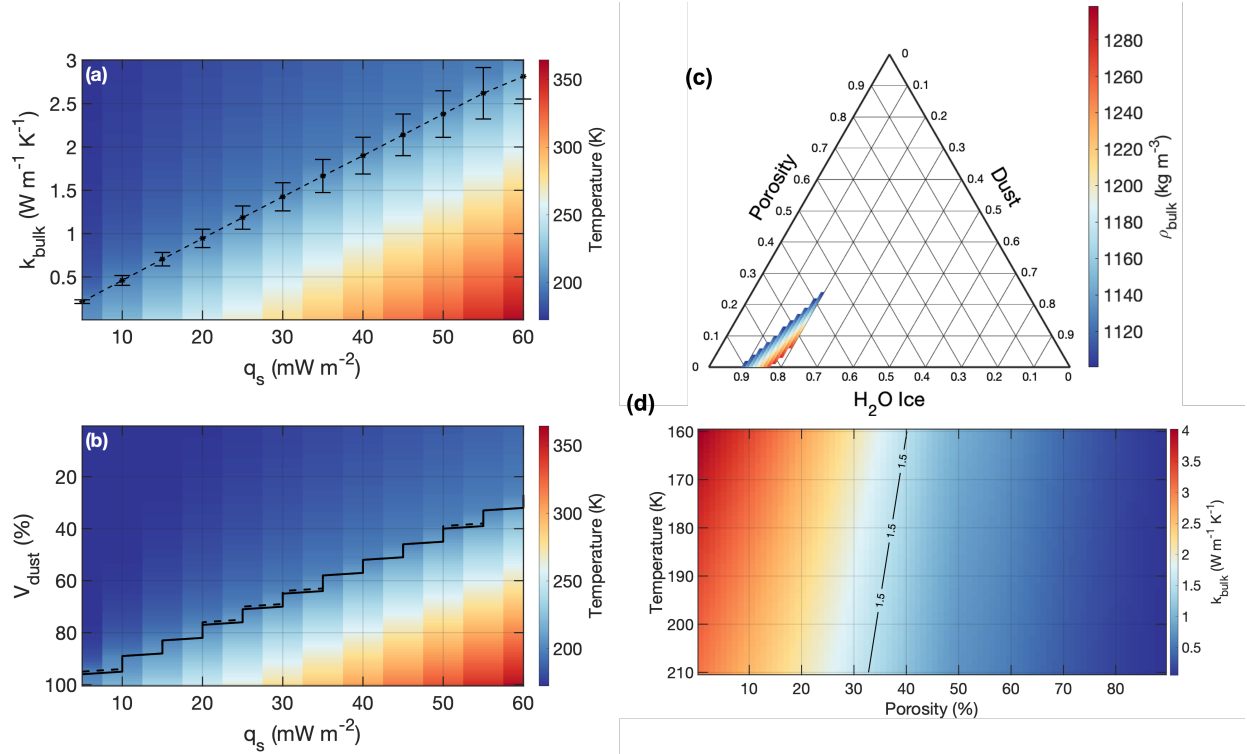

**Figure 1. (a)** Bulk thermal conductivity of the SPLD necessary for basal melting ($T_m = 199$ K) as a function of $q_s$. $k_{bulk}$ lower than denoted by the black line would lead to basal melting. The error bars show the negligible effect of $T_s$ (between 165-170 K) on the $k_{bulk}$ required for basal melting. The color shows the basal surface temperature at 1.5 km depth. **(b)** Similar to (a), but showing the volume of dust required for basal melting as a function of the surface heat flow. The effect of dust's thermal conductivity variation is negligible (notice overlap of solid and dotted line values of 0.015 and 0.039 W m⁻¹ K⁻¹). **(c)** Ternary diagram showing the possible mixture of pore space, ice, and dust that can satisfy the bulk density estimate of the SPLD. **(d)** Thermal conductivity of ice as a function of pore-space and temperature using the mixing model of Calonne et al. (2019). The black line marks the minimum porosity needed for $k_{bulk} \leq 1.5$ W m⁻¹ K⁻¹.

## 3.2. Can a thin layer of impurity at the base of the SPLD enable basal melting?

We now consider if a layer of dust-rich ice (hereafter referred to as 'dirty ice') at the base of the SPLD can aid in basal melting (Fig. 2 a). In this scenario, we vary the thickness (h) and the volumetric dust content ($\theta$) of the dirty ice layer. The SPLD above the dirty ice is assumed to have 20% dust volumetrically. Figure 2 (b) shows an example of the thermal evolution of a 1.5 km thick ice sheet, with the bottom 300 m composed of dirty ice ($\theta = 95$ %), and a surface heat flow of 30 mW m⁻². In this extreme scenario, the temperature within the region occupied by dirty ice increases over time, and basal melting is observed at the base (region under the white contour). We perform a parameter sweep to constrain the minimum h and $\theta$ required under the SPLD for basal melting to occur (Fig. 2 c, d). Basal melting is only observed if the dirty ice layer contains more than 90 % dust volumetrically, which is substantially more than the dust/ice ratio found even in the nucleus of comets [*Küppers et al.*, 2005]. Furthermore, the dirty ice layer's thickness must be in excess of 100 meters for basal melting to occur. If basal melting does occur, it is much more likely to occur at the base, where dirty ice is in direct contact with the crust (Fig. 2 b). A reasonable variation in $k_{dust}$ and $T_s$ has a negligible effect on the feasibility of basal melting in this scenario (Fig. 2 c, d).

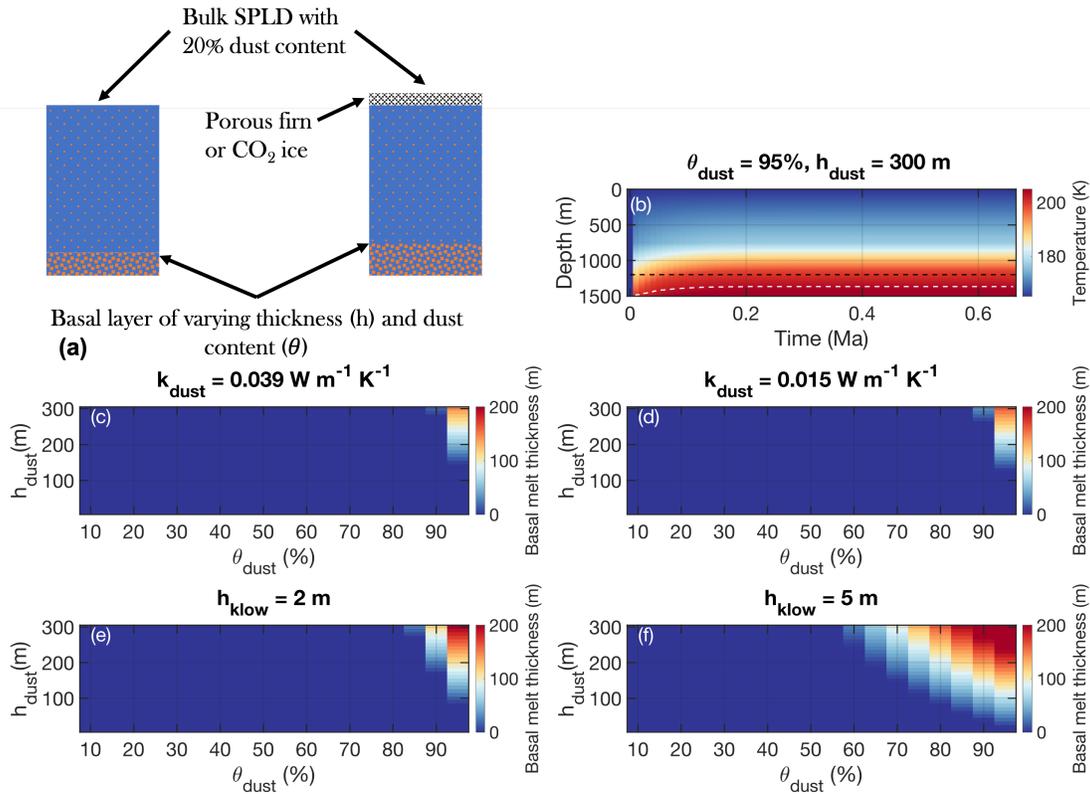

**Figure 2.** (a) Schematic of the case where we consider if dirty ice of varying thickness and dust content at the base of the SPLD can enable basal melting. (b) An example of a temperature distribution of the SPLD as a function of depth and time for a surface heat flow of 30 mW m$^{-2}$ assuming a 1.2 km thick ice sheet, superposed on a 300 m thick dirty-ice with 95% dust content. The black line shows the 300 m mark from the base, and the white contours show the region under which basal melting occurs. (c) Result from a parameter sweep that shows the minimum thickness and dust content of dirty ice layer required for basal melting. The color plot shows the thickness of the resulting melt. (d) Same as (c) but assuming a $k_{dust}$ of 0.015 W m$^{-1}$ K$^{-1}$. (e) Similar to (c) and (d) but showing the basal melt thickness when a 2-meter thick low conductivity layer at the top of the SPLD is considered. (f) Similar to (e), but assuming a 5-meter thick low conductivity layer at the top of the SPLD.

The thermal inertia of the Martian south polar region is relatively low, suggestive of the presence of a low thermal conductivity unit such as firn or $CO_2$ in the upper few meters [*Putzig et al.*, 2005]. The thermal conductivity of even the lowest density porous ice lies around 0.06 W·m$^{-1}$·K$^{-1}$ [*Calonne et al.*, 2019], whereas $CO_2$ ice can have a much lower density of 0.02 W·m$^{-1}$·K$^{-1}$ [*Sori and Bramson*, 2019]. We now assume the presence of $CO_2$ ice on top of the SPLD and vary its thickness to assess the effect of a low conductive layer on the feasibility of basal melting. We find that a low thermal conductivity unit like $CO_2$ ice can significantly aid basal melting when a dirty ice layer is present at the base of the SPLD (Fig. 2); however, even when a 5-meter low conductive layer is assumed to be present, basal melting only occurs with dirty ice containing more than 50% dust volumetrically. An unrealistic aspect of this model is that the thermal conductivity of the dust at the base of the SPLD would be equivalent to the bulk-thermal conductivity of the dust measured at the surface. The $k_{dust}$ values reported by InSight and laboratory measurements of the Mars simulant are for the bulk, porous dust at the surface [*Grott et al.*, 2021; *Yu et al.*, 2022]. The overburden pressure and pore-filling ice would reduce the pore space within the dust at the base of the SPLD; thus, the results presented in Figure 2 are for a highly optimistic scenario. Furthermore, if a thick layer of dirty ice does exist at the base of the SPLD, then the resultant melt may resemble a sludge or hydrated smectite clays [*Smith et al.*, 2021] instead of a subglacial lake.

### 3.4. What would it require to create a 20 - 30 km wide lake under the SPLD?

A recent magmatic intrusion underneath the SPLD can augment the near-surface temperature and enable basal melting. An intrusion of variable width is intruded into the crust at various depth; the heat from the intrusion raises the near surface temperature of the SPLD while at the same time the temperature of the intrusion declines due to the conduction of heat to the surrounding cold country rocks (Supplementary Movie 1; Supplementary Movie 2). The cooling rate depends on the overall geometry of the intrusion and the temperature of the surrounding rock (Fig. 3 b). In general, larger intruded bodies at depth cool much slower due to a higher degree of insulation than smaller bodies at shallow depths (Fig. 3 b). The heat from the magmatic intrusion also affects the near-surface temperature of the crust and the base of the SPLD, possibly leading to basal melting. Figure 3 c – f shows the resultant width of the basal melt that can form from an intrusion of a given width and depth over time.

In these models, we set the temperature of the crust by assuming a surface heat flow of 30 mW m$^{-2}$ (Text. S3; Fig. S5). While there are many parameters at play, the key takeaway from this exercise is that intrusions need to be relatively large and close to the surface of Mars to induce any degree of basal melting. For example, a narrow dyke-like intrusion (W = 1 - 3 km), regardless of its proximity to the surface, would not be able to provide enough heat for a 20 – 30 km wide subglacial lake as proposed [*Lauro et al.*, 2021]. In general, an intrusion must have a similar width to the width of the subglacial lake.

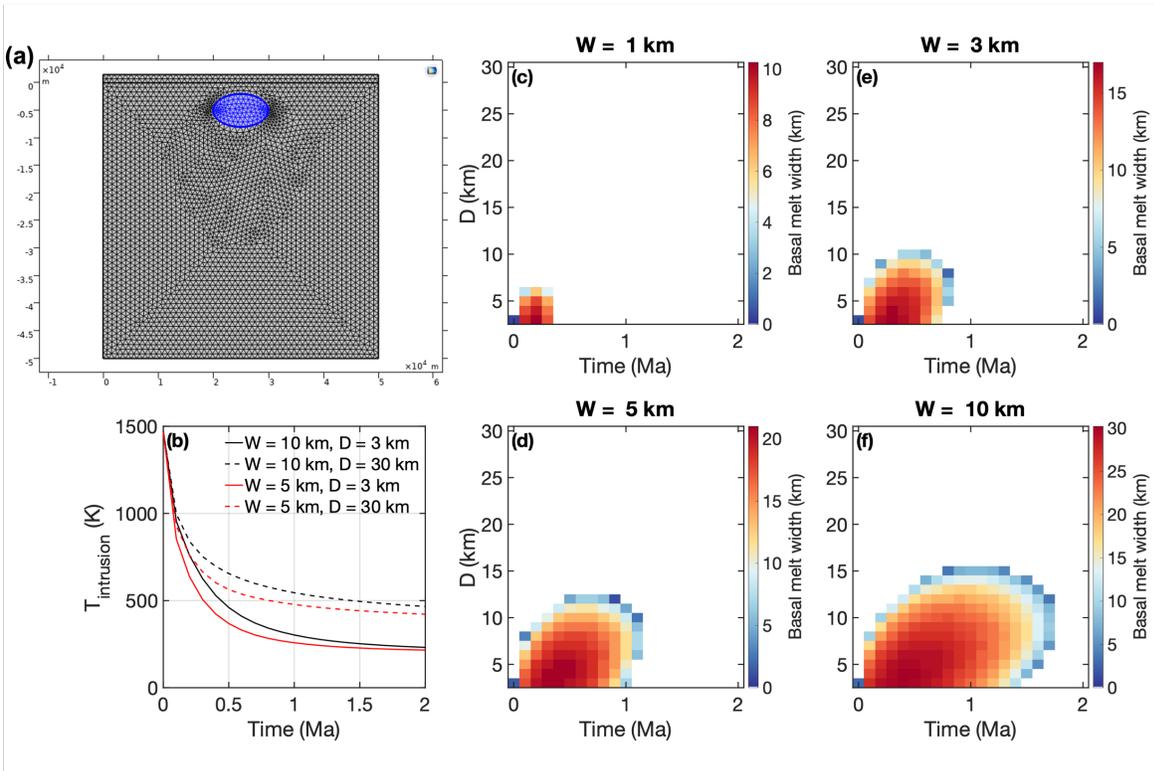

**Figure 3. (a)** An example of a mesh showing the magmatic intrusion (in blue) of width W at depth D within the Martian crust of thickness 50 km. We vary W and D in this experiment to ascertain their values that would enable basal melting of a 20-30 km wide lake under the SPLD. **(b)** The cooling rate of the magmatic intrusion of various W and D within the Martian crust as a function of time. **(c) – (f)** Results from parameter sweep show the width of basal melt as a function of time for magmatic intrusions of various W and D. Only large magmatic intrusions (c and d) at shallow depths can form 20-30 km wide subglacial lakes under the SPLD.

Currently, there is no way to assess if such magmatic intrusions may be present at depth under the SPLD. To date, no major Bouguer gravity anomalies or crustal magnetic field anomalies, possibly related to an intrusion in the south polar region have been identified. Figure 3 c – f also highlights that any subglacial lakes formed by magmatic intrusion are metastable over geological timescales. As the temperature of the intruded body cools down, so does the near-surface temperature, refreezing any subglacial lakes. Even if their presence is confirmed, the transient nature of intrusion sustained subglacial lakes, as illustrated here, has obscure implications for their astrobiological significance.

### 3.5. Feasibility of basal melting with lower salinity of the SPLD

One of the outstanding issues with the possibility of briny subglacial lakes in the SPLD is the amount of salt required to sustain such lakes. The notion that the purported lake exists underneath the SPLD due to a high concentration of perchlorates with a eutectic temperature of 199 K is entirely speculative; no evidence currently supports that notion. Further, given the reported dimension of the main putative lake (20 × 30 km) [*Lauro et al.*, 2022] and the minimum molal concentration of salt required for supercooled brine [*Toner et al.*, 2014], it has previously been shown that salt masses over $10^{12}$ kg would be required to form such a voluminous brine solution under the SPLD [*Ojha et al.*, 2021]. The perchlorate salts on Mars are derived globally from the atmospheric oxidation of HCl [*Catling et al.*, 2010; *Wilson et al.*, 2016]; thus, the process responsible for such significant sequestration of perchlorate at the SPLD remains unexplained. Given the difficulty reconciling such a large volume of perchlorate salts within the SPLD, we also explore the thermophysical conditions required for forming subglacial lakes with higher melting temperatures. Figure S6 shows that if the melting temperature were slightly higher than the assumed 199 K value, basal melting would be implausible with a $q_b$ of 30 mW m$^{-2}$ for most saline ice. For example, the basal temperature of the SPLD would not exceed the eutectic temperature of Ca(ClO$_3$)$_2$ and Mg(ClO$_3$)$_2$ [*Hanley et al.*, 2012] even assuming SPLD with 50 % dust. Even if the surface temperature of the SPLD were miraculously augmented to 180 K, basal melting of ice with reasonable salinity would be unlikely (Fig. S6).

## 4  Conclusions

We consider several scenarios for potentially forming a subglacial lake in the SPLD. First, we show that basal melting is unlikely to occur with available heat flow constraints and reasonable assumptions about the presence of pore space and impurities like dust and CO$_2$. Second, a dirty ice layer at the base of the SPLD can potentially enable basal melting; however, the dirty ice layer must be extremely dust rich and thick. Melt in this scenario would occur at the base resembling wet soil overlain by dry dirty ice instead of a lake. Third, a recent magmatic intrusion can augment the near-surface temperature and enable basal melting; however, a relatively large intrusion at shallow depths would be required. Magmatic intrusion-enabled subglacial lakes would also be relatively short-lived with limited astrobiological potential. Finally, basal melting of other chlorate and perchlorate species is even more unlikely, even with a 50 % dust content in the SPLD. In summary, exceptional circumstances would be necessary to form subglacial lakes in the SPLD.

## Acknowledgments

LO is supported by a startup grant from Rutgers University.

## Open Research

No new data is generated as part of this work. COMSOL model outputs are deposited on Zenodo and can be accessed at the following location https://doi.org/10.5281/zenodo.7658862 (CCC).

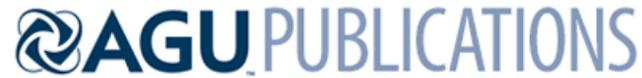

*Geophysical Research Letters*

Supporting Information for

**Thermophysical assessment on the feasibility of basal melting in the south polar region of Mars**

**Lujendra Ojha[1], Jacob Buffo[2], Baptiste Journaux[3]**

[1]Department of Earth and Planetary Sciences, Rutgers University, Piscataway, NJ, USA.
[2]Thayer School of Engineering. Dartmouth College. Hanover, NH, USA.
[3]Department of Earth and Space Science, University of Washington, Seattle, WA, USA.

Corresponding author: Lujendra Ojha (luju.ojha@rutgers.edu)

**Contents of this file**

**Text S1. Minimum timescale for the thermal evolution models**

The response time to thermal perturbations via conduction only is given by the thermal time constant ($\tau$):

$$\tau = \frac{l^2}{\pi\kappa};$$

<div align="right">(eqn S1)</div>

where $l$ is the length scale over which the thermal perturbation occurs and $\kappa$ is the thermal diffusivity. For an ice thickness of 1.5 km, the thermal time constant assuming $\kappa = 10^{-6}$ m² s⁻¹ is of order $10^5$ years. Thus, we run our thermal models for a minimum of 2 million years.

### Text S2. Comparison of the COMSOL output to analytical solutions

The Stefan problem is a classical heat transfer problem that involves finding the position of the phase change boundary between a homogeneous solid and its melt ($x_m$) as a function of time ($t$). For a system where the solid phase is initially at its melting temperature ($T_0$) and is subjected to a constant temperature amplification ($T_1 > T_0$) at one boundary, the temperature in the liquid half-space $T(x, t)$ and the evolving position of the melting front ($x_m(t)$) are given by:

$$T(x,t) = T_1 - (T_1 - T_0) \frac{\operatorname{erf}\left(\frac{x}{2\sqrt{\kappa t}}\right)}{\operatorname{erf}(\lambda)}, \text{ for } 0 \leqslant x \leqslant x_m(t),$$

$$x_m(t) = 2\lambda\sqrt{\kappa t},$$

$$\lambda \exp(\lambda^2) \operatorname{erf}(\lambda) = \frac{St}{\sqrt{\pi}}.$$

<div align="right">(eqn S2)</div>

Here, erf is the error function, and $\kappa = k/\rho c$ is the thermal diffusivity of the liquid. The coefficient '$\lambda$' depends on the Stefan number ($St$) which is given by:

$$St = c(T_1 - T_0)/L_f;$$

<div align="right">(eqn S3)</div>

Where $L_f$ is the latent heat of the solid-liquid phase change and c is the specific heat capacity of the liquid. For simplicity, we set St = 3, and $L_f = 1$ J kg⁻¹, which gives a value of 0.9138 for $\lambda$ and 0.2727 J kg⁻¹ K⁻¹ for c. All other parameters are listed in the Table S1.

Figure S1 shows a comparison of the evolving temperature distribution in the liquid half-space during a 200 second simulation. A good agreement is found between the analytical solution and the COMSOL phase change simulation result, guaranteeing the accuracy of the apparent heat capacity formulation implemented throughout the manuscript. A similar validation of the COMSOL results has been provided by [*Ogoh and Groulx*, 2010].

### Text S3. Details on COMSOL modeling of basal melting from magmatic instrusion

We first create a mesh 51.5 km thick and 100 km wide (Fig. S4). The upper 1.5 km of the mesh is represented by the SPLD superposed on a 50 km thick crust. The initial steady-state thermal profile of the crust is modeled via the following:

$$T = \frac{\rho H z}{\kappa_c}\left(T_{cr} - \frac{z}{2}\right) + \frac{q_b z}{\kappa_c} \qquad for\ z < T_{cr}$$

$$T = \frac{\rho H T_{cr}^2}{2\kappa_m} + \frac{q_b z}{\kappa_m} \qquad for\ z > T_{cr}.$$

<div align="right">(eqn S3)</div>

where $\kappa$ is thermal conductivity of the crust (3 W m⁻¹ K⁻¹), T is temperature (K), z is depth (m), $\rho$ is the density of the crust (3000 kg m⁻³). H is the heat production rate per unit mass, $\kappa_m$ is the thermal conductivity of the mantle (2.5 W m⁻¹ K⁻¹), and $q_b$ is the basal heat flux. The heat production rate and $q_b$ are varied such that the surface heat flow (i.e., $q_s = q_r + q_b$) in the south polar region of Mars does not exceed 30 mW m⁻² as shown in Fig. S5. The net heat contribution by crust vs. mantle plays a notable role in the temperature profile of a planet's lithosphere; however, the near-surface temperature is not strongly modulated, as shown by geotherms in Fig. S5.

A coarse mesh represents the crust, whereas regions close to and above the magmatic intrusion, including the SPLD, are represented by extremely fine mesh with 500 elements and an element ratio of 8 (Fig. S4). We impose an initial condition of Ts = 165 K on top of and within the SPLD and set the temperature of the magmatic intrusion to 1500 K. The thermal profile of the crust changes with time due to the surface heat flow augmentation by the magmatic intrusion, and we run the models for a maximum of 2 million years (see Text S1). The thermal conductivity, specific heat capacity, and density of the SPLD is modeled using SeaFreeze and equation 5-6 in the main text (also see Table S2). The left and right sides of the mesh are thermally insulated.

The effective thermal conductivity gets modified with phase changes, so we utilize COMSOL's 'Conjugate Heat Transfer' interface, which combines Heat Transfer and Single-Phase Flow interfaces to describe heat transfer in solids and fluids and nonisothermal flow in fluids. The SPLD is characterized by a phase-change material within the model with a melting temperature of 199 K. At the interface between the SPLD and the Martian crust, we impose a 'no slip' boundary condition to prevent any fluid from leaving the domain.

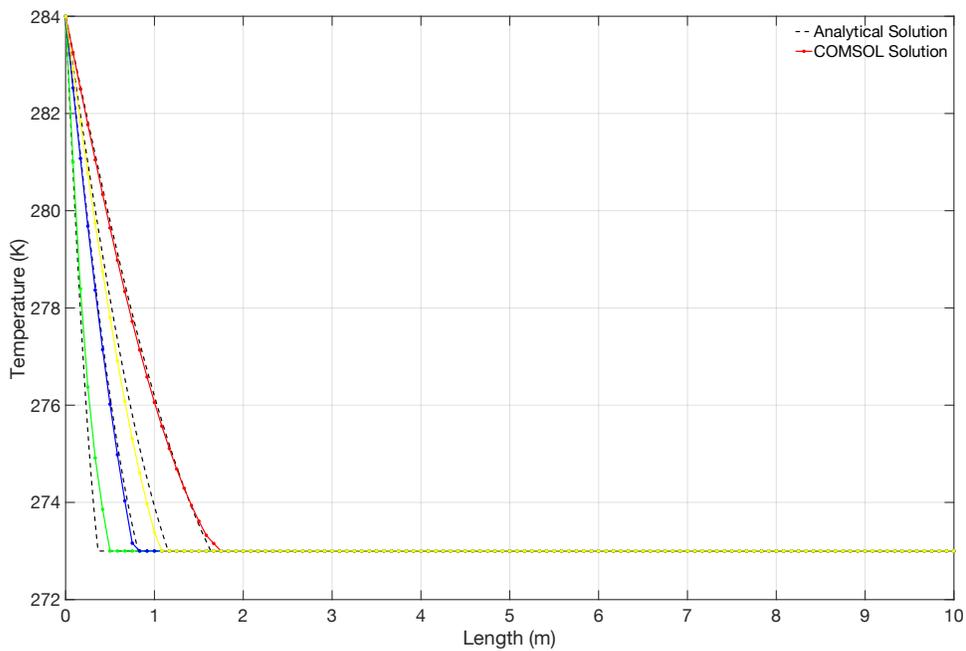

**Figure S1.** A comparison of the COMSOL output and analytical solution to the Stefan problem described in Text S2 at 10 seconds (green), 50 seconds (blue), 100 seconds (yellow), and 200 seconds (red).

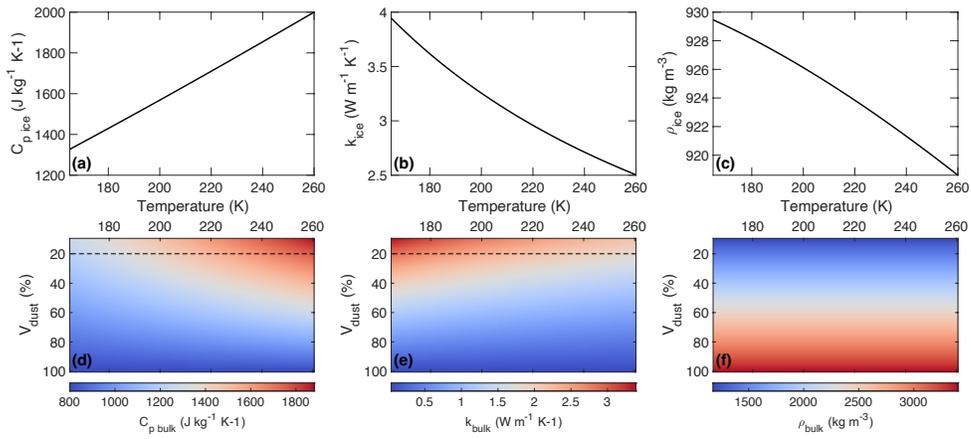

**Figure S2.** An example of temperature-dependent profiles of specific heat ($C_{p\,ice}$) **(a)**, thermal conductivity ($k_{ice}$) **(b)**, and density ($\rho_{ice}$) **(c)** of pure-water ice. These initial profiles are coupled with a volumetric mixing model to estimate the $C_{p\,bulk}$ **(d)**, $k_{ice}$ **(e)**, and $\rho_{bulk}$ of water ice and dust mixture of varying proportions.

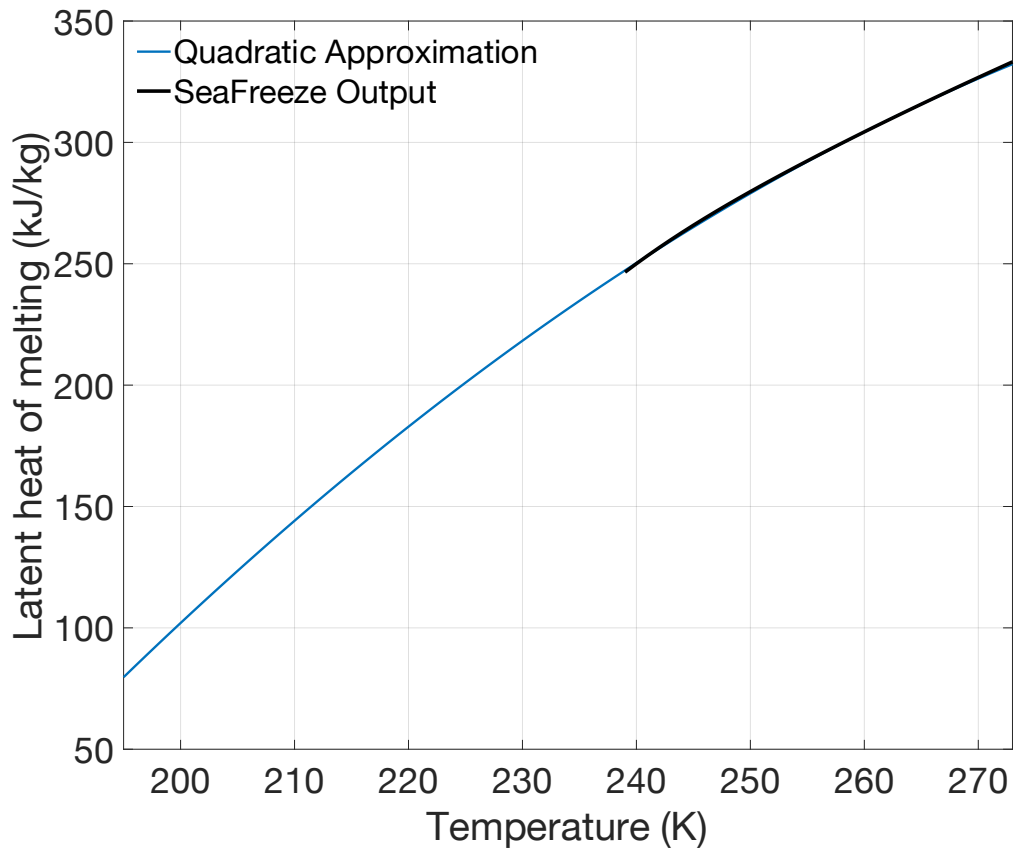

**Figure S3.** Temperature dependent latent heat of melting of water ice. The output from SeaFreeze up to 239 K is shown in black; quadratic extrapolation down to 195 K shown in blue.

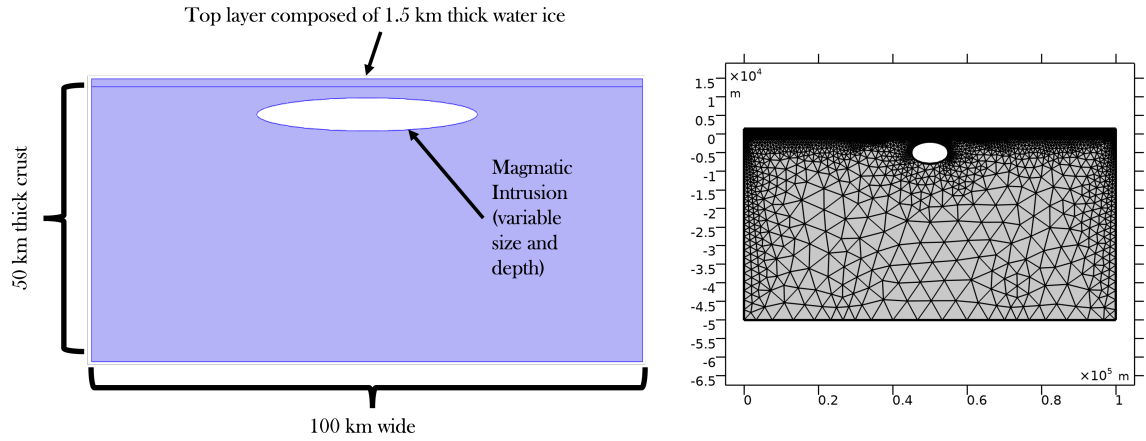

**Figure S4.** Schematic and the mesh used in the 2D study of a basal melting via magmatic intrusion.

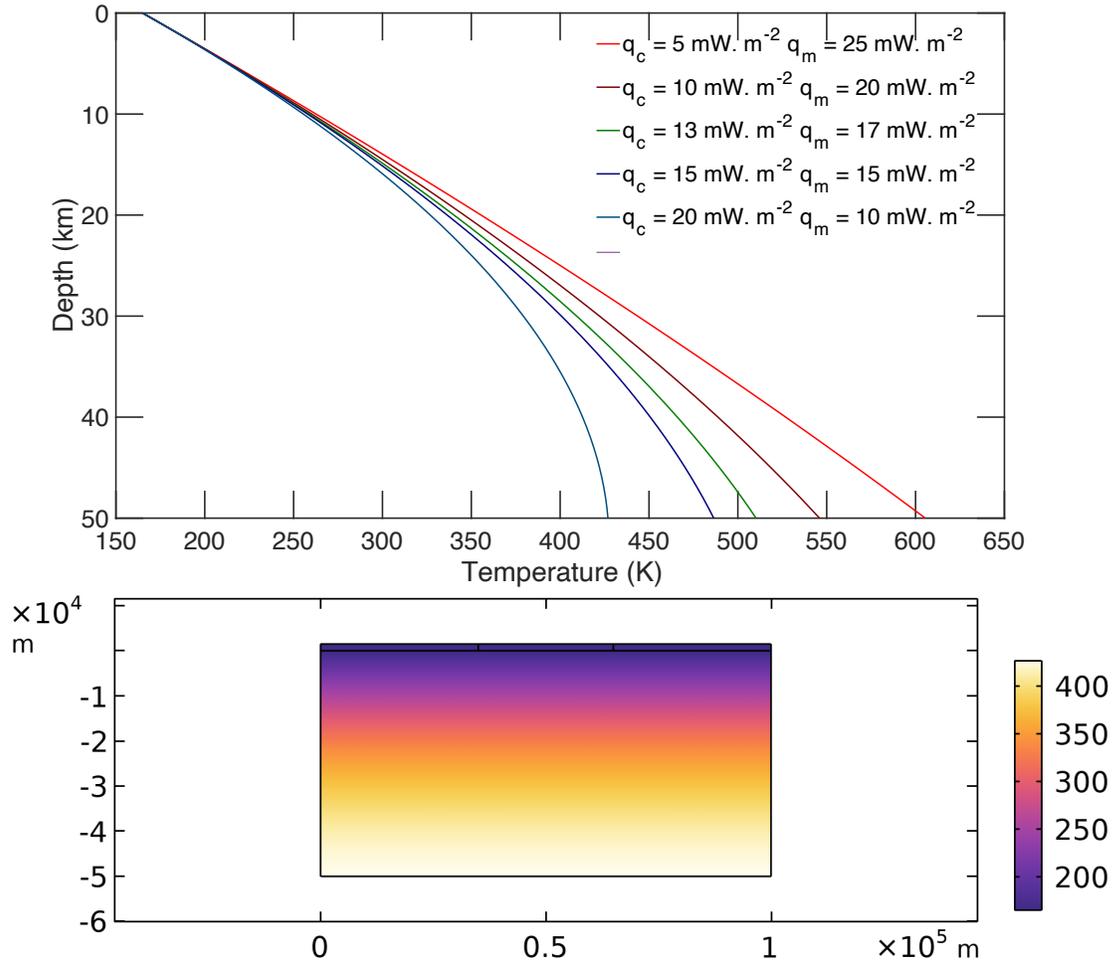

**Figure S5.** **(Top)** Steady state geotherm of the Martian crust assuming variable $q_c$ and $q_m$. See Table S2 for values of thermophysical parameters of the crust. **(Bottom)** Initial temperature profile of the crust in 2D in COMSOL before the introduction of a high temperature magmatic intrusion.

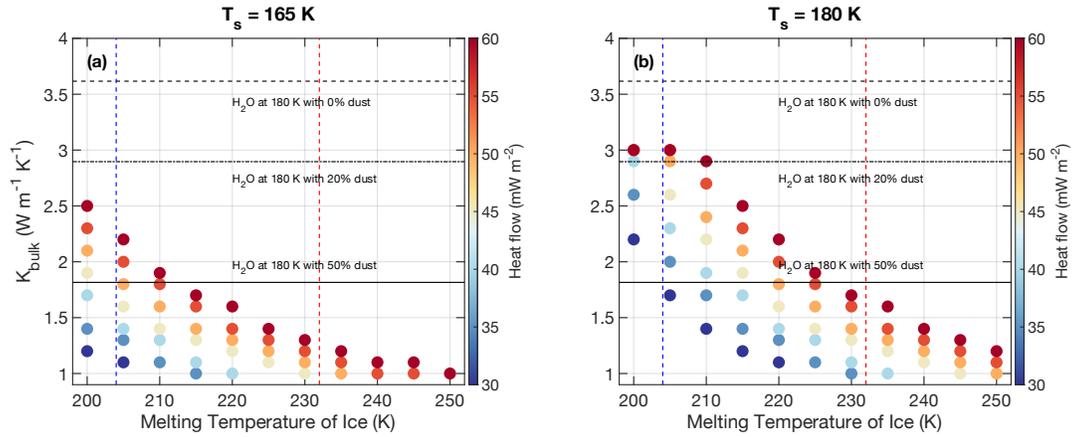

**Figure S6.** Heat flow required for basal melting as a function of the melting temperature and the $k_{bulk}$ of the SPLD for $T_s$ of 165 **(a)** and 180 K **(b)**. The three horizontal lines show the thermal conductivity of pore-free ice at 180 K with various proportion of dust. The blue and red vertical lines mark the eutectic temperatures for $Ca(ClO_3)_2$ and $Mg(ClO_3)_2$ of 232 K and 204 K, respectively.

| Parameters | Value [units] |
| --- | --- |
| T1 | 284[K] |
| T0 | 273[K] |
| x | (0:0.01:10) [m] |
| t | (0:0.1:200) [s] |
| $\kappa$ | 0.0040 $(m^2\ s^{-1})$ |
| $\rho$ | 914 $[kg\ m^{-3}]$ |
| St | 3 [unitless] |
| c | 0.2727 [J kg⁻¹ K⁻¹] |
| K | 1 [W m⁻¹ K⁻¹] |
| λ | 0.9138 [unitless] |

**Table S1.** *Parameters and their values used in assessing the accuracy of the apparent heat capacity method of COMSOL via comparison to the analytical solution of the Stefan problem.*

| Parameters | Value |
|---|---|
| Surface Temperature of SPLD (T$_s$) | 165 K [*Sori and Bramson*, 2019] |
| k$_{ice}$ | Temperature dependent; SeaFreeze derived [*Journaux et al.*, 2020] |
| Cp$_{ice}$ | Temperature dependent; SeaFreeze derived [*Journaux et al.*, 2020] |
| ρ$_{ice}$ | Temperature dependent; SeaFreeze derived [*Journaux et al.*, 2020] |
| k$_{dust}$ | 0.015 [*Yu et al.*, 2022]  – 0.039 [*Grott et al.*, 2021] W m⁻¹ K⁻¹ |
| Cp$_{dust}$ | 800 J·kg⁻¹·K⁻¹ |
| ρ$_{dust}$ | 2500 - 3000 kg m⁻³ |
| k$_{CO2}$ | 0.02 W m⁻¹ K⁻¹ [*Sori and Bramson*, 2019] |
| k$_{crust}$ | 3 W m⁻¹ K⁻¹ [*Turcotte and Schubert*, 2014] |
| T$_{intrusion}$ | 1500 K [*Arndt et al.*, 2008] |
| L | SeaFreeze derived  [*Journaux et al.*, 2020] |

**Table S2.** *Parameters and their values used in assessing the feasibility of basal melting in the SPLD.*